\documentstyle[11pt,epsf]{article}

\newcommand\Appendix[1]{\par
\setcounter{section}{0}
 \setcounter{equation}{0}
 \renewcommand{\thesection}{Appendix \Alph{section}}
\section{#1}
 \def\theequation{\Alph{section}.\arabic{equation}}}
\newcommand\appendixn[1]{\par
 \setcounter{equation}{0}
 \renewcommand{\thesection}{Appendix \Alph{section}}
\section{#1}
 \def\theequation{\Alph{section}.\arabic{equation}}}

\newcommand{\cZ}{{\cal Z}}
\newcommand{\cS}{{\cal S}}

\newcommand{\cD}{{\cal D}}
\newcommand{\Z}{{Z \!\!\! Z}}
\newcommand{\eq}[1]{(\ref{#1})}
\newcommand{\dual}{\mbox{}^{\ast}}

\newcommand{\CK}[1]{\mbox{\scriptsize c}_{\mbox{$\scriptstyle #1$}}}
\newcommand{\beq}{\begin{equation}}
\newcommand{\eeq}{\end{equation}}
\newcommand{\beqn}{\begin{eqnarray}}
\newcommand{\eeqn}{\end{eqnarray}}
\newcommand{\nsum}[2]{\sum_{ #1(\CK{#2}) \in \Z }}

\newcommand{\nddsum}[2]{\sum_{\stackrel{\scriptstyle \dual #1(\dual\CK{#2})
\in \Z} {\delta \dual #1=0}}}
\newcommand{\intpi}{\int\limits_{-\pi}^{+\pi} {\cD}}
\newcommand{\intinf}{\int\limits_{-\infty}^{+\infty} {\cD}}
\newcommand{\dd}{\mbox{d}}
\newcommand{\expb}[1]{\exp\left\{ #1 \right\} }

\date{}
\begin{document}
\rightline{\normalsize KANAZAWA 96-18}
\rightline{\normalsize September 1996}
\vspace{1cm}
\centerline{\normalsize\bf EFFECTIVE CONSTRAINT POTENTIAL FOR}
\baselineskip=22pt
\centerline{\normalsize\bf ABELIAN MONOPOLE IN $SU(2)$ LATTICE GAUGE
THEORY}
\vspace*{0.6cm}
\centerline{\footnotesize M.N.~CHERNODUB${}^1$, M.I.~POLIKARPOV${}^{1,2}$
and A.I.~VESELOV${}^1$}
\baselineskip=13pt
\vspace*{0.3cm}
\centerline{\it ${}^1$ITEP, B.Cheremushkinskaya 25, Moscow,
117259, Russia}
\centerline{\it ${}^2$Department of Physics, Kanazawa
University, Kanazawa 920-11, Japan}
\vspace*{0.8cm}

\abstract{
We describe numerical calculation results for the probability
distribution of the value of the monopole creation operator in the
$SU(2)$ lattice gluodynamics. We work in the maximal abelian
projection. It occurs that at low temperatures, below the
deconfinement phase transition, the maximum of the distribution is
shifted from zero, which means that the effective constraint potential
is of the Higgs type. Above the phase transition the minimum of the
potential (the maximum of the monopole field distribution) is at the
zero value of the monopole field. This fact confirms the existence of
the abelian monopole condensate in the confinement phase of lattice
gluodynamics, and agrees with the dual superconductor model of the
confining vacuum.}

\vspace*{0.6cm}
\normalsize\baselineskip=15pt
\setcounter{footnote}{0}
\renewcommand{\thefootnote}{\alph{footnote}}

\section{Introduction}

The monopole mechanism of the color confinement
\cite{Man76,tHo,KrScWi87} is generally accepted by the lattice
community. Still there are many open questions \cite{Gre96}. In the
lattice gluodynamics it is very important to find the order parameter,
constructed from the monopole field for the deconfinement phase
transition. The first candidate is the value of the monopole
condensate, which should be nonzero in the confinement phase and
vanish at the phase transition. To study the monopole condensate, we
need an explicit expression for the operator $\Phi_{mon} (x)$, which
creates the abelian monopole at the point $x$. The operator
$\Phi_{mon}(x)$, found for the compact electrodynamics with the
Villain form of the action by Fr\"ohlich and Marchetti \cite{FrMa87},
was studied numerically in \cite{Wiese}. In Section~2 we construct the
analogous monopole creation operator for an arbitrary abelian
projection of the lattice $SU(2)$ gluodynamics. The numerical results
presented in Section 3 are obtained for the maximal abelian
projection. As shown by many numerical simulations for this projection
the gluodynamic vacuum behaves as the dual superconductor (see reviews
\cite{Suz93,Pol96} and references therein). In \cite{DiGi}
another form of the monopole creation operator was studied; it is
shown that its expectation value vanishes in the deconfinement
phase. The monopole creation operator, constructed in Section~2 is
positive, and therefore its expectation value cannot
vanish at $T = T_c$. Still the results of the numerical studies of the
effective potential for $\Phi_{mon}$ presented in Section 3, clearly
indicate that the monopole condensate exists in the confinement phase
of lattice gluodynamics and does not exist in the deconfinement
phase. The analogous claim is done in ref.\cite{IvPoPo93}, where the
monopole condensate is calculated on the basis of the percolation
properties of the monopole currents. The monopole creation operator in
the monopole current representation is studied in
ref.\cite{Nak96}. Again, the monopole creation operator depends on the
temperature as the disorder parameter.

In Appendices A and B we prove that the operator used in the numerical
calculations create the monopole in the lattice $U(1)$ theory with the
general form of the action. Appendix C contains the brief description
of the differential form notations on the lattice, these notations are
used in Appendices A and B.

\section{Monopole Creation Operator}

First we give a formal construction of the monopole creation operator
for the abelian projection of the $SU(2)$ gluodynamics.  Let us
parametrize the $SU(2)$ link matrix in the standard way:
$U^{11}_{x\mu} = \cos \phi_{x\mu}\, e^{i\theta_{x\mu}}; \
U^{12}_{x\mu} = \sin
\phi_{x\mu}\, e^{i\chi_{x\mu}};$ $\ U^{22}_{x\mu} = U^{11 *}_{x\mu}; \
U^{21}_{x\mu} = - U^{12 *}_{x\mu};$ $\ 0 \le \phi \le \pi/2, \ -\pi <
\theta,\chi \le \pi$.
The plaquette action in terms of the angles $\phi, \ \theta $ and
$\chi$ can be written as follows:

\beq
S_P  =  \frac{1}{2}\mbox{Tr}\, U_1 U_2 U_3^+ U_4^+ = S^a + S^n + S^i\;,
\label{SP}
\eeq
where

\beq
S^a  = \cos \theta_P\,
\cos\phi_1\, \cos\phi_2\, \cos\phi_3\, \cos\phi_4,
\label{Sa} \\
\eeq
$S^n$ and $S^i$ describe the interaction of the fields $\theta$ and $\chi$
and selfinteraction of the field $\chi$~\cite{ChPoVe95};
$\theta_P  = \theta_1 + \theta_2 - \theta_3 - \theta_4$,
here the subscripts $1,...,4$ correspond to the links of the plaquette:  $1
\rightarrow \{x,x+\hat{\mu}\},...,4 \rightarrow \{x,x+\hat{\nu}$\}.
For a fixed abelian projection,
each term $S^a$, $S^n$ and $S^i$ is invariant under the residual
$U(1)$ gauge transformations:

\beqn
    \theta_{x\mu} & \to & \theta_{x\mu} +\alpha_x -\alpha_{x+\hat{\mu}}
     \label{u1th}\;,\\
    \chi_{x\mu} & \to & \chi_{x\mu} +\alpha_x + \alpha_{x+\hat{\mu}}\;.      \label{u1chi}
\eeqn

We define the operator which creates the monopoles at the point $x$ of
the dual lattice as follows:

\beqn
   \Phi_{mon}(x) = \exp \left\{ \beta [ - S(\theta_P,...) + S(\theta_P +
   W_P(x),...)] \right\}\,,  \label{Ux1}
\eeqn
the function $W_P(x)$ being defined by eq. \eq{Phi}.
Substituting (\ref{SP}--\ref{Sa}) into \eq{Ux1}, we get

\beqn
   \Phi_{mon}(x) = \exp \left\{ \sum_P \tilde\beta \left[ - \cos(\theta_P) +
   \cos(\theta_P + W_P(x)) \right] \right\}\,, \label{Ux2}
\eeqn
where
$\tilde\beta = \cos\phi_1 \cos\phi_2 \cos\phi_3\cos\phi_4 \,\beta$.
Effectively the monopole creation operator shifts all abelian
plaquette angles $\theta_P$.

For the compact electrodynamics with the Villain type of the action
the above definition coincides with the definition of Fr\"ohlich and
Marchetti \cite{FrMa87}. For the general type of the action in compact
electrodynamics we can use the described above construction. The proof
is outlined in Appendices A,B. The gluodynamics in the abelian
projection contains the compact gauge field $\theta$ and the charged
{\it vector} field $\chi$. The action \eq{SP} in terms of the fields
$\theta$ and $\chi$ is rather nontrivial, and at the moment we can not
prove that the above construction of the monopole creation operator is
valid in this case. However, there is a proof for a similar Abelian --
Higgs model, with the general type of the action, and this proof is
analogous to one given in Appendix B. Moreover the numerical results,
presented in the next section, clearly show that the introduced
operator is the order parameter for the deconfinement phase
transition.

\section{Numerical Results}

The numerical results are obtained on the lattice $4\cdot L^3$, for $L
= 8,10,12,14,16$. The preliminary numerical results are published in
ref. \cite{ChPoVe96}.  We extrapolate our results to the infinite
volume since near the phase transition the finite volume effects are
very strong. We impose anti--periodic boundary conditions in space
directions for the abelian fields, since the construction of the
operator $\Phi_{mon}$ can be done only in the time slice with the
anti--periodic boundary conditions. Periodic boundary conditions are
forbidden due to the Gauss law: we input a magnetic charge into the
finite box.  Formally, equation \eq{divD} for $\dual D$ admits no
solution in the finite box with periodic boundary conditions. To
impose anti--periodic boundary conditions on the abelian fields the
$C$--periodic boundary conditions should be imposed on the nonabelian
gauge fields \cite{Wie92}. In the case of $SU(2)$ gauge group the
$C$--periodic boundary conditions are almost trivial: on the boundary
we have $U_{x,\mu} \to
\Omega^+ U_{x,\mu} \Omega,\,\,\Omega = i \sigma_2$.
To get the order parameter for the deconfinement phase transition we
study the probability distribution of the operator $\Phi_{mon}$;
we calculate the expectation value
$<\delta (\Phi - \Phi_{mon}(x))>$. The effective constraint potential,

\beq
V_{eff}(\Phi) =
-\ln (<\delta (\Phi - \frac{1}{V}\sum_x \Phi_{mon}(x))>)
\eeq
has more physical meaning than the probability distribution. The
calculation of $V_{eff}(\Phi)$ is time consuming, and we
present our results for $V(\Phi)$, defined as follows:

\beq
V(\Phi) = -ln(<\delta (\Phi - \Phi_{mon}(x))>).
\eeq
In Figs. 1(a) and 1(b) $V(\Phi)$ is shown for the confinement and the
deconfinement phases, the calculations being performed on the lattice
$4\cdot12^3$.

In the confinement phase the minimum of $V(\Phi)$ is shifted from
zero, while in the deconfinement phase the minimum is at the zero
value of the monopole field $\Phi$. We have used the positive
operator, $\Phi_{mon}(x) > 0$ \eq{Ux2}, however in the dual
representation the creation operator of the monopole
\eq{divD} is not positively definite: the sign is lost when we perform
the inverse duality transformation (cf. eq. \eq{divD} with
eq. \eq{Phi}). The potential shown in Fig.~1(a) corresponds to the
Higgs type potential. The value of the monopole field, $\Phi_c$, at
which the potential has a minimum is equal to the value of the
monopole condensate. The potential shown in Fig.~1(b) corresponds to
the trivial potential with a minimum at the zero value of the field:
$\Phi_c = 0$. The dependence of the minimum of the potential,
$\Phi_c$, on the spatial size of the lattice, $L$, is shown in Fig.~2.
The gauge fields are generated by the standard heat bath method. At
each value of $\beta$ 2000 update sweeps are performed to thermalize
the system. The maximal abelian projection \cite{KrScWi87} corresponds
to the maximization of the quantity $R = \sum_{x,\mu}
Tr(U_{x\mu}\sigma_3 U^{+}_{x\mu}\sigma_3)$. For our relatively small
lattices the overrelaxation algorithm \cite{MaOg90,HKM91} and simple
local maximization method \cite{KrScWi87} give approximately the same
results, and we use the local maximization method. We stop our gauge
fixing sweeps when $Z=10^{-5}$, here $Z$ \cite{HKM91} is the lattice
analogue of the quantity $<|(\partial_\mu + i A^3_\mu)A_\mu^+|^2>$
which should be zero if the maximal abelian projection is fixed
exactly. We checked that the more precise gauge fixing do not change
the results for $\Phi_c$ inside the statistical errors.  For each
value of $\beta$ at the lattice of definite size we use 100 gauge
field configurations separated by 300 Monte Carlo sweeps. For each
configuration we calculated the value of the monopole creation
operator at 20 randomly chosen lattice points.  Therefore for each
value of $\beta$ at the lattice of definite size we have 2000 values
of $\Phi_{mon}$. We use these values to calculate the quantity
$\Phi_c$ (the maximum of the probability distribution
$\rho(\Phi_{mon})$ is at $\Phi_{mon} =
\Phi_c$).

We fitted the data for $\Phi_c$ by the formula $\Phi_c = A L^\alpha +
\Phi_c^{inf}$, where $A$, $\alpha$ and $\Phi_c^{inf}$ are the
fitting parameters. It occurs that $\alpha = -1$ within statistical
errors. Fig.~3 shows the dependence on $\beta$ of the value of the
monopole condensate, extrapolated to the infinite spatial volume,
$\Phi_c^{inf}$. It is clearly seen that $\Phi_c^{inf}$ vanishes at
the point of the phase transition and it plays the role of the order
parameter.

\vspace*{0.8cm}
\noindent
{\normalsize\bf Acknowledgments}\par\vspace*{0.4cm}

The authors are grateful to P.~van Baal, A.~Di~Giacomo, R.~Haymaker,
T.~Ivanenko, Y.~Matsubara, A.~Pochinsky, T.~Suzuki and U.~-J.~Wiese
for useful discussions. M.I.P. feels much obliged for the kind
reception given to him by the staff of Theoretical Physics Institute, the
University of Minneapolis, and the staff of the Department of Physics of the
Kanazawa University, where the part of this work has been done. This work
was supported by the JSPS Program on Japan -- FSU scientists collaboration,
by the Grant INTAS-94-0840, and by the grant number 96-02-17230a, financed
by the Russian Foundation for Fundamental Sciences.

\Appendix{ }

Here we construct the monopole creation operator for the compact
electrodynamics with the general type of the action. A similar
construction exists for the compact Abelian--Higgs model with the
general type of the action. First we perform the duality
transformation of the partition function for the $4D$ lattice compact
electrodynamics with the general type of the action $\cS (\dd\theta +
2 \pi) = \cS (\dd\theta)$,

\beq
 {\cZ} = \intpi \theta
         \expb{ -\cS (\dd\theta)}, \label{ZQED}
\eeq
We use the notations of the calculus of differential forms on the
lattice \cite{BeJo82} (see also Appendix C). The symbol
$\int\cD\theta$ denotes the integral over all link variables $\theta$.
First consider the Fourier series for the Boltzmann factor:

\beq
     {\cZ} = const. \intpi \theta \nsum{n}{2}
     F(n) e^{i (n,\dd \theta) },\,\,
        F(n) = \intpi X \expb{- \cS [X]
        -i (n, X)}\,. \label{FT}
\eeq
Integrating over $\theta$ we get the partition function of the dual theory:

\beq
{\cZ}^d = \nsum{k}{2}
\expb{- \dual \cS (\dd \dual k)},
\eeq
where $n=\delta k$, $\dual \cS(n)= - \ln F(n)$. We can represent
$\cZ$ as the following limit of the partition
function for the Abelian--Higgs theory:

\beqn
{\cZ}^d = \lim_{\eta \rightarrow\infty}
\intpi \dual \varphi\intinf \dual B
\sum_{ \scriptstyle {\dual k(\CK{1}) \in \Z}}
\exp\bigl\{-\dual \cS(\dd \dual B/2\pi) -
\eta \|\dual B - \dd\dual\varphi + 2 \pi \dual k\|^2 \bigr\},
\label{AH2}
\eeqn
here $\dual S(\dd \dual B/2\pi)$ is the kinetic energy of the dual
gauge field $\dual B$ (the analogue of ${\tilde F}_{\mu\nu}^2$) and
the Higgs field $\exp\{i\, \dual\varphi\}$ carries magnetic charge, since
it interacts via the covariant derivative with the dual gauge field
$\dual B$. The Dirac operator \cite{Dir55},

\beq
{\Phi_{mon}}^d(x) = e^{i\dual\varphi}\cdot
\expb{-i(\dual D_x,\dual B)},\,\,
\delta \dual D_x = \dual \delta_x \label{divD}
\eeq
is the gauge invariant monopole creation operator. It creates the
cloud of photons and the monopole at the point $x$. In \eq{divD}
$\dual \delta_x$ is the lattice $\delta$--function, it equals to unity
at the site $x$ of the dual lattice and is zero at the other sites. Note
that in the above formulas the radial part of the Higgs field which
carries the magnetic charge is fixed to unity.

Coming back to the original partition function \eq{ZQED} we get the
expectation value of the monopole creation operator in terms of the
fields $\theta$:

\beq
<\Phi_{mon}(x)> =\frac{1}{\cZ} \intpi\theta
         \expb{ -\cS(\dd\theta  + W_P)},\,\,
W_P = 2 \pi \delta\Delta^{-1}(D_x-\omega_x)), \label{Phi}
\eeq
where the Dirac string attached to the monopole \cite{FrMa87}, is
represented by the integer valued 1-form $\dual \omega_x$, which
satisfies the equation: $\delta \dual \omega_x = \dual \delta_x$.

\appendixn{ }

Here we represent the partition function \eq{ZQED} for the
compact electrodynamics with the general type of the action as the sum
over the monopole currents. First we insert the unity
$1 = \intinf  G \, \delta(G-n)$ into the sum (\ref{FT})
and use the Poisson summation formula:
$\sum_{n} \delta(G-n) = \sum_{n} e^{2 \pi i (G, n)}$. We get:

\beq
     \cZ = const \cdot
     \intpi \theta  \intinf  G \nsum{n}{2}
     F(G)\, \expb{i (\dd  \theta + 2 \pi n, G)}\,. \label{poisson}
\eeq
Here $G$ is a real--valued two--form. It is possible to change the
summation variable $n$: ${\displaystyle \sum_n f(n) = \sum_q
\sum_{\delta \dual j = 0} f(m[j] + \dd q)}$, where $n = m[j] + \dd q\,,
\,\, \dd m[j] = j\,,\,\, \dd j =0$. Now we change the compact
integration variable, $\theta$, to
the noncompact one $A$: ${\displaystyle \sum_n}\intpi  \theta\,
f(\dd \theta + 2\pi n) = {\displaystyle \sum_{\delta \dual j = 0}}
\intinf A \, f(\dd A + 2\pi \delta\Delta^{-1} j)$, where $A = \theta +
2 \pi \Delta^{-1}\delta m[j] + 2\pi q $ and
we use the Hodge--de--Rahm formula: $m = \delta {\Delta}^{-1} j +
\dd {\Delta}^{-1} \delta m$. The integral over $A$ gives the constraint
$\delta(\delta G)$, which we solve introducing the new integration
variable $H$, $G = \delta H$. Taking into account
the relation $\dd \delta {\Delta}^{-1} j =
j$, valid for any $j$, such that $\dd j = 0$,
we finally get the representation of the
partition function as a sum over the conserved monopole currents:

\beq
        \cZ = const \cdot\nddsum{j}{1} e^{- \cS_{mon}(\dual j)}\,.
        \label{ZBKT}
\eeq
where

\beq
        \cS_{mon}(\dual j) = - \ln \left(\,\, \intinf H F(\delta H)
        \expb{2 \pi i (\dual H,\dual j) } \right)\,.
        \label{Smon}
\eeq
The monopole action is nonlocal due to the integral over $H$. If we
start from the Villain action $\cS^V(\dd \theta) = - ln \sum_n
\expb{-\beta \|\dd\theta + 2\pi n \|^2}$, then the integral over
$H$ in \eq{Smon} is Gaussian, and we get the well known expression
\cite{BaMyKo77} for the monopole action: $\cS_{mon}^V(\dual j) = 4
\pi^2 \beta (\dual j,
\Delta^{-1} \dual j)$. Using transformations similar
to \eq{poisson}--\eq{ZBKT} for the expectation value of the monopole
creation operator \eq{Ux1}, we get

\beq
<\Phi_{mon}(x)> =\frac{1}{\cZ}\sum_{\delta\dual j = \delta_x}
\expb{\cS_{mon}(\dual j - \dual D_x)}\, ,
\eeq
where $D_x$ is defined by eq. \eq{divD}.
Therefore $\Phi_{mon}(x)$ creates a non--closed monopole world
trajectory starting at the point $x$, which shows that, indeed,
$\Phi_{mon}(x)$ is a monopole creation operator.

\appendixn{ }

Let us briefly summarize the main notions from the theory of
differential forms on the lattice \cite{BeJo82}.  The advantage of the
calculus of differential forms consists in the general character of
the expressions obtained. Most of the transformations depend neither
on the space--time dimension, nor on the rank of the fields. With
minor modifications, the transformations are valid for lattices of any
form (triangular, hypercubic, random, {\it etc}). A differential form
of rank $k$ on the lattice is a function $\phi_{k}$ defined on
$k$-dimensional cells $c_k$ of the lattice, {\it e.g.}, the scalar
(gauge) field is a 0--form (1--form). The exterior differential
operator {\it d} is defined as follows:

\beq
(\dd \phi ) (c_{k+1}) =\sum_{\CK{k} \in \partial\CK{k+1}} \phi(c_{k}).
\label{def-dd}
\eeq
Here $\partial c_{k}$ is the oriented boundary of the $k$-cell
$c_{k}$. Thus the operator {\it d} increases the rank of the form by unity;
$\dd \varphi$ is the link
variable constructed, as usual, in terms of the site angles $\varphi$, and
$\dd A$ is the plaquette variable constructed from the link variables $A$.
The scalar product is defined in the standard way:
if $\varphi$ and $\psi$ are $k$-forms, then
$(\varphi,\psi)=\sum_{c_k}\varphi(c_k)\psi(c_k)$, where $\sum_{c_k}$ is the
sum over all cells $c_k$.
To any $k$--form on the $D$--dimensional lattice there
corresponds a $(D-k)$--form $\dual\Phi(\dual c_k)$ on the dual lattice,
$\dual c_k$ being the $(D-k)$--dimensional cell on the dual lattice. The
co-differential $\delta=\dual \dd \dual$ satisfies the partial
integration rule: $(\varphi,\delta\psi)=(\dd\varphi,\psi)$.
Note that $\delta \Phi(c_k)$ is a $(k-1)$--form and
$\delta \Phi(c_0) = 0$. The norm is defined by: $\|a\|^2=(a,a)$; therefore,
$\|B - \dd\varphi+2\pi n\|^2$ in \eq{AH2} implies summation over all links.
$\nsum{l}{1}$ denotes the sum over all configurations of the integers $l$
attached to the links $c_1$. The action \eq{AH2} is invariant under the
gauge transformations $B' = B + \dd \alpha$, $\varphi' = \varphi + \alpha$
due to the well known property $\dd^2 = \delta^2 = 0$. The
lattice Laplacian is defined by: $\Delta = \dd\delta + \delta\dd$.

\newpage
\begin{figure}[htb]

\vspace{-5.cm}
\centerline{\epsfxsize=0.55\textwidth\epsfbox{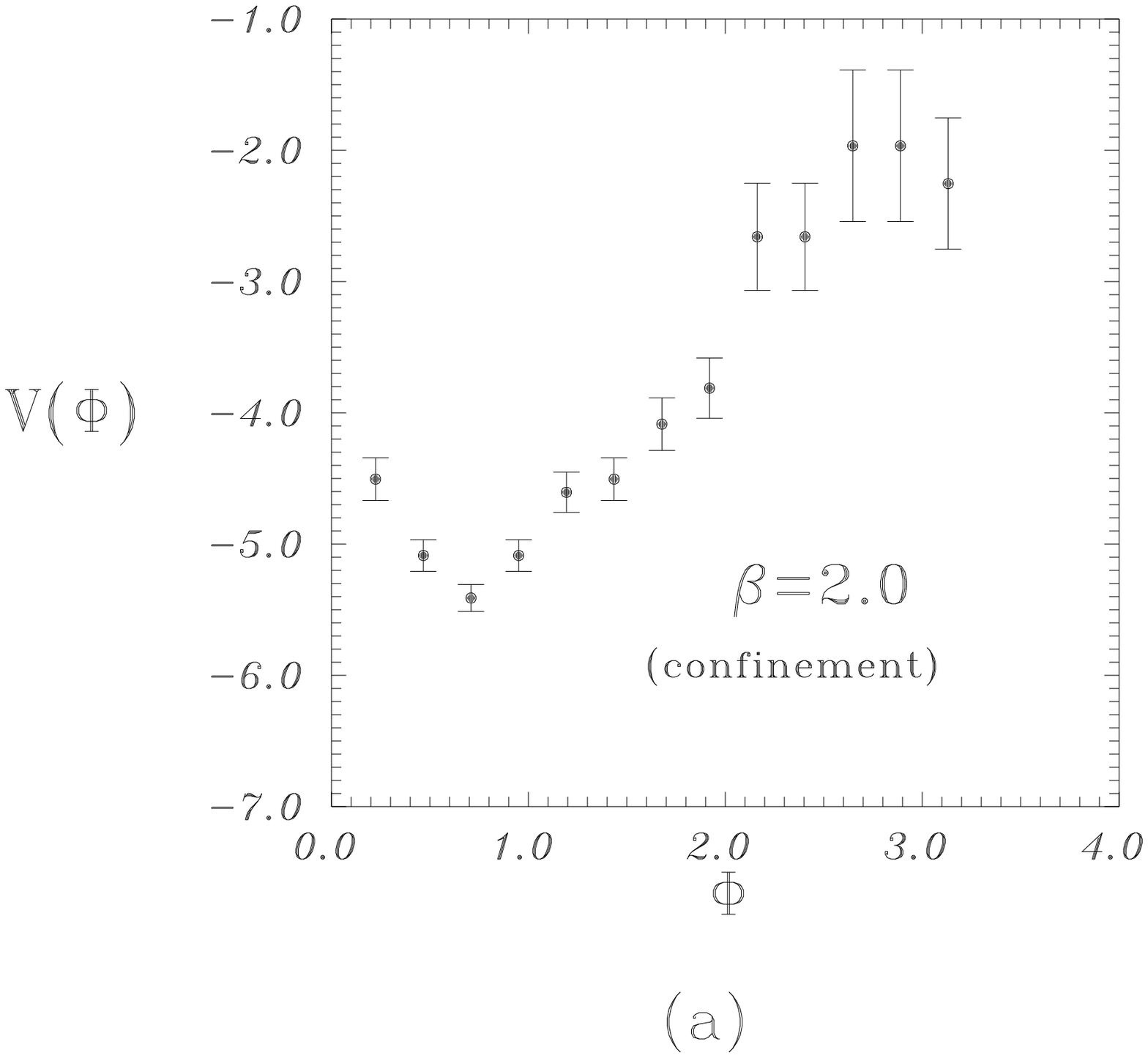}} 
\vspace{-2.9cm}
\centerline{\epsfxsize=0.55\textwidth\epsfbox{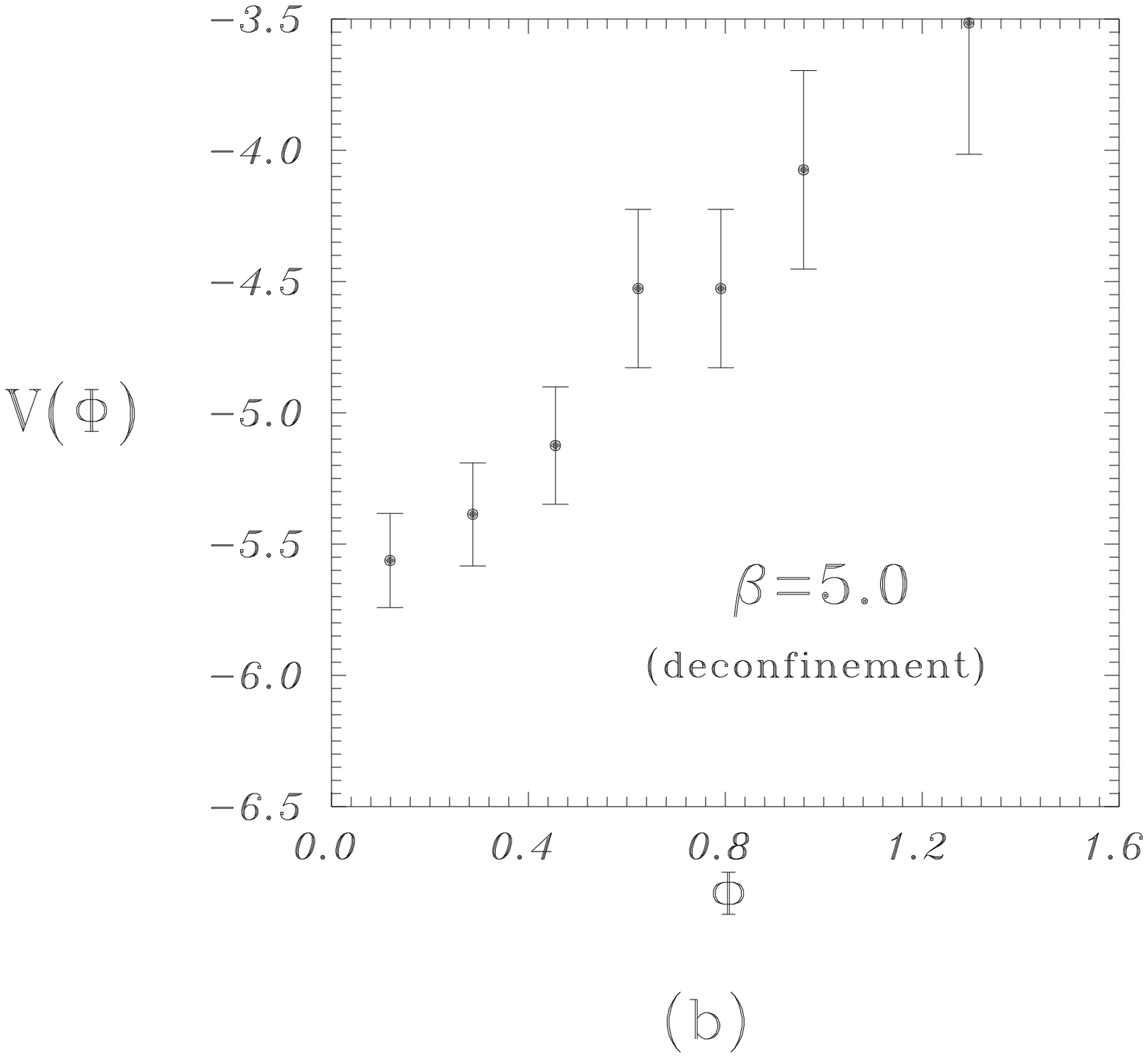}} 
\vspace{1.5cm}
\caption{
$V(\Phi)$ for the confinement~(a) and the deconfinement~(b) phases.}
\vspace{-0.6cm}
\end{figure}

\newpage

\begin{figure}[htb]

\vspace{-5.cm}
\centerline{\epsfxsize=.65\textwidth\epsfbox{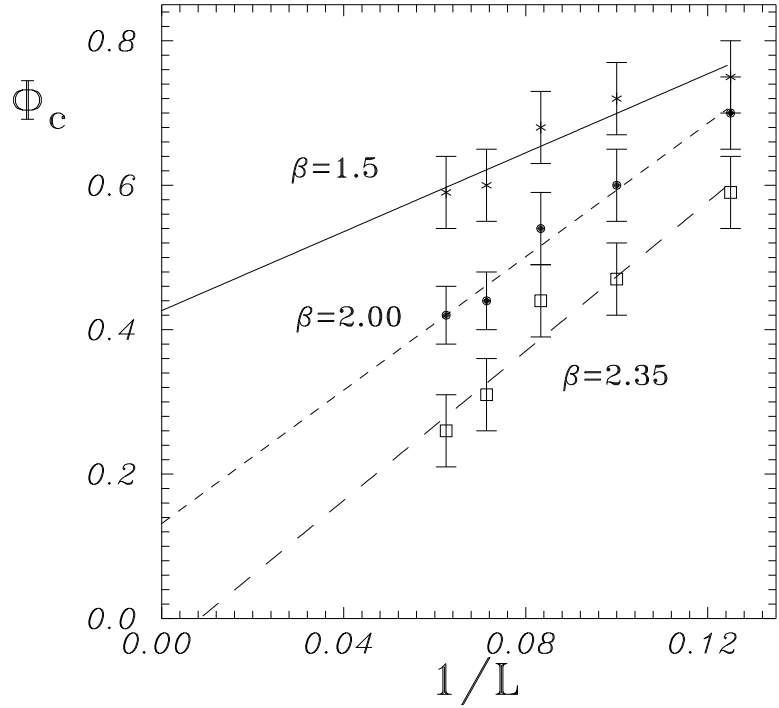}} 
\vspace{0.1cm}
\caption{
The dependence of $\Phi_c$ on the spatial size of the lattice for
three values of $\beta$.}
\vspace{-0.6cm}
\end{figure}
\begin{figure}[htb]
\vspace{-5.cm}
\centerline{\epsfxsize=.65\textwidth\epsfbox{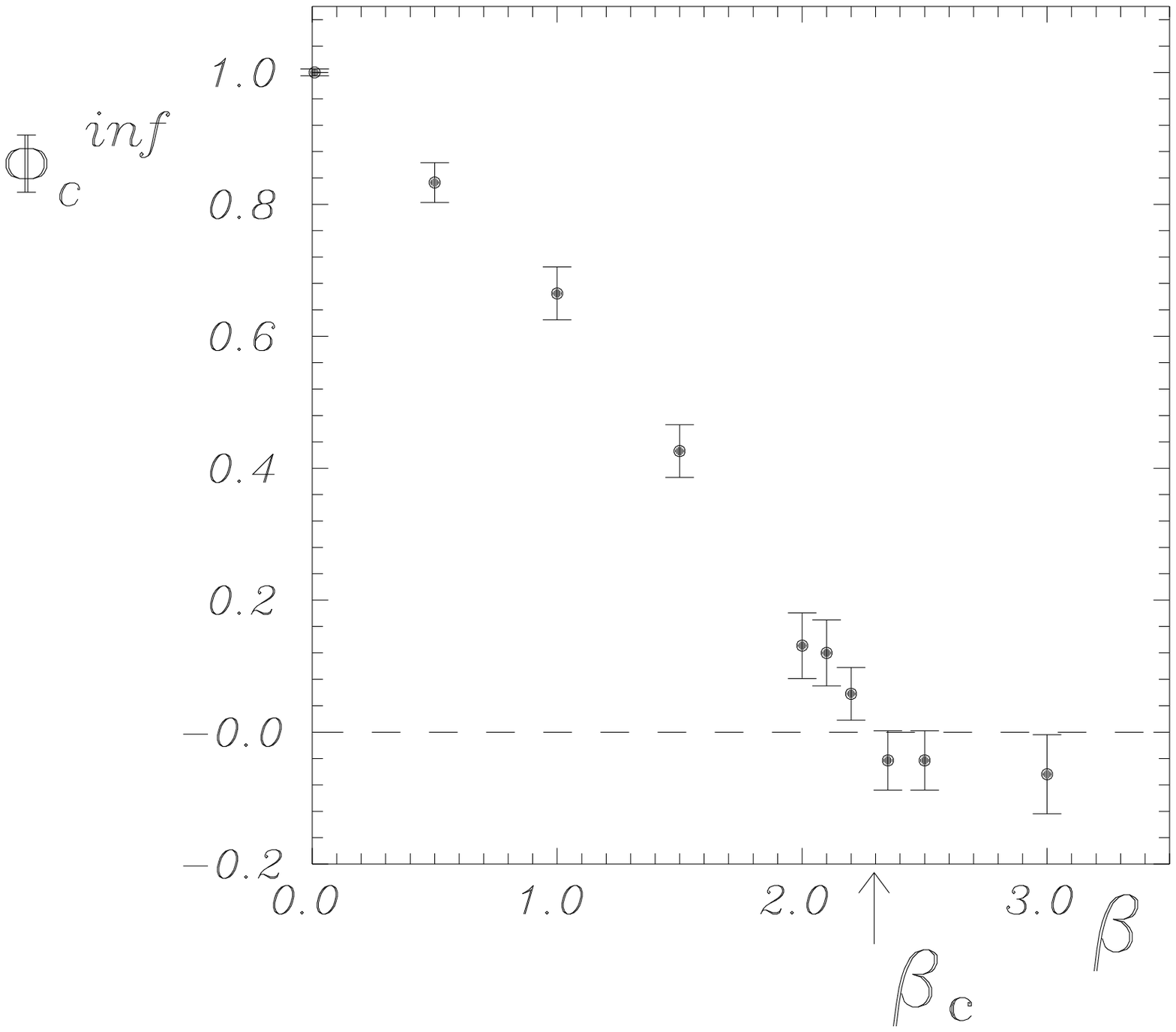}} 
\vspace{0.1cm}
\caption{
The dependence of $\Phi_c^{inf}$ on $\beta$.}
\vspace{-0.6cm}
\end{figure}


\begin{thebibliography}{10}

\bibitem{Man76} S.~Mandelstam, {\it Phys.~Rep.}, 23C (1976) 245.

\bibitem{tHo} G.~{'t~Hooft}, "High Energy Physics",
{\rm {Z}ichichi, Editrice Compositori, Bolognia}, 1976;\\
G.~'t~Hooft,  {\it Nucl.~Phys.}, {\bf B190} [FS3] (1981) 455.

\bibitem{KrScWi87} A.S.~Kronfeld, M.L.~Laursen, G.~Schierholz and
U.J.~Wiese, Phys.Lett.198B
(1987) 516;\\
A.S.~Kronfeld, G.~Schierholz and U.J. Wiese, Nucl.Phys.
B293 (1987) 461.

\bibitem{Gre96}  L.~Del~Debbio, M.~Faber, J.~Greensite and S.~Olejnik,
talk given at Lattice 96 Symposium; {\tt hep-lat/9607053}.

\bibitem{FrMa87}
J.~{Fr\"{o}hlich} and P.A.~Marchetti, {\it Commun.~Math.~Phys.},
{\bf 112} (1987) 343.

\bibitem{Wiese}
L.~Polley and U.J.~Wiese, {\it Nucl.~Phys.} {\bf B356} (1991) 629;\\
M.I.~Polikarpov, L.~Polley and U.J.~Wiese,
{\it Phys.~Lett.} {\bf B253} (1991) 212.

\bibitem{Suz93}
T.~Suzuki, {\it Nucl.~Phys.}, {\bf B} (Proc.~Suppl.) {\bf 30} (1993) 176.

\bibitem{Pol96} M.I.~Polikarpov, preprint KANAZAWA-96-19, talk given at
Lattice 96 Symposium, {\tt hep-lat/9609020}

\bibitem{DiGi} L.~Del~Debbio, A.~Di~Giacomo and G.~Paffuti,
{\it Phys.~Lett.} {\bf B349} (1995) 513;\\
L.~Del~Debbio, A.~Di~Giacomo, G.~Paffuti and P.~Pieri,
{\it Phys.~Lett.} {\bf B355} (1995) 255.

\bibitem{IvPoPo93}
T.L.~Ivanenko, A.V.~Pochinsky and M.I.~Polikarpov,
{\it Phys.~Lett.} {\bf 302B} (1993) 458.

\bibitem{Nak96} N.~Nakamura, V.~Bornyakov, S.~Ejiri,
S.~Kitahara, Y.~Matsubara and T.~Suzuki, preprint KANAZAWA-96-16;
{\tt hep-lat/9608004}.

\bibitem{ChPoVe95}
M.N.~Chernodub, M.I.~Polikarpov and A.I.~Veselov,
{\it Phys.~Lett.} {\bf B342} (1995) 303.

\bibitem{ChPoVe96}
M.N.~Chernodub, M.I.~Polikarpov and A.I.~Veselov,
{\it Nucl.~Phys.} B (Proc. Suppl.) {\bf 49} (1996) 307,
{\tt hep-lat/9512030}.

\bibitem{Wie92}
U.J. Wiese, {\it Nucl.Phys.} {\bf B375} (1992) 45;\\
A.S. Kronfeld and U.J. Wiese, {\it Nucl.~Phys.} {\bf B401} (1993) 190.

\bibitem{Dir55}
P.A.M.~Dirac, {\it Canad.~J.~Phys.}, {\bf 33} (1955) 650.

\bibitem{BaMyKo77} T.~Banks, R.~Myerson and J.~Kogut, {\it Nucl.~Phys.},
{\bf B129} (1977) 493.

\bibitem{MaOg90} J.E.~Mandula and M.~Ogilivie,
{\it Phys.~Lett.} {\bf B248} (1990) 156.

\bibitem{HKM91} S.~Hioki, S.~Kitakhara, O.~Miyamura, S.~Ohno and 
T.~Suzuki, {\it Phys.~Lett.} {\bf B271} (1991) 201.

\bibitem{BeJo82}
P.~Becher and H.~Joos, {\it Z.~Phys.}, {\bf C15} (1982) 343.

\end{thebibliography}
\end{document}